\documentclass{ifacconf}

\usepackage{amssymb}
\usepackage{mathtools}
\usepackage{eurosym}
\usepackage{xcolor}
\usepackage{graphicx}
\usepackage{tikz}
\usepackage{dsfont}
\usepackage{enumitem}
\usepackage{pgfplots}
\usepackage{stackengine}

\usetikzlibrary{intersections, pgfplots.fillbetween}

\newcommand{\be}{\begin{equation}}
	\newcommand{\ee}{\end{equation}}

\newcommand{\dd}{\,\mathrm{d}} 
\newcommand{\N}{\mathcal{N}}

\newcommand{\R}{\mathbb{R}} 
\newcommand{\A}{\mathcal{A}}  
\newcommand{\mc}{\mathcal}

\renewcommand{\S}{\mathcal{S}} 

\usepackage{graphicx}      
\usepackage{natbib}        
\begin{document}
\begin{frontmatter}

\title{Nash equilibria of the pay-as-bid auction
	with K-Lipschitz supply functions} 


\author[First]{Martina Vanelli} 
\author[First]{Giacomo Como} 
\author[First]{Fabio Fagnani}

\address[First]{Department of Mathematical Sciences, Politecnico di Torino
	(e-mail: {martina.vanelli,giacomo.como, fabio.fagnani}@polito.it).}

\begin{abstract}  
		We model a system of $n$ asymmetric firms selling a homogeneous good in a common market through a pay-as-bid auction. Every producer chooses as its strategy a supply function returning the quantity $S(p)$ that it is willing to sell at a minimum unit price $p$. The market clears at the price at which the aggregate demand intersects the total supply and firms are paid the bid prices. We study a game theoretic model of competition among such firms and focus on its equilibria (Supply function equilibrium). The game we consider is a generalization of both models where firms can either set a fixed quantity (Cournot model) or set a fixed price (Bertrand model).  
		Our main result is to prove existence and provide a characterization of (pure strategy) Nash equilibria in the space of $K$-Lipschitz supply functions.
\end{abstract}
 

\end{frontmatter}
\section{Introduction}
The progressive liberalization of electricity markets 
motivates the need to develop realistic and robust models for the analysis of the strategic bidding problem 
(\cite{ventosa}, \cite{roozbehani2010stability}, \cite{tang2016model}, \cite{paccagnan2016aggregative}).
Pricing rules in oligopolistic wholesale electricity auctions are mainly two: the uniform price rule and the pay-as-bid rule (\cite{rassenti}, \cite{fabra}). In a uniform price auction, electricity is paid/sold at the market-clearing price, regardless of the offers that bidders actually made. On the other hand, in the pay-as-bid auction (also called discriminatory price auction), the remuneration is the bid price. 

From a game-theoretic point of view (\cite{maryam}), appropriate models for studying wholesale markets for electricity  are Supply Function Equilibrium (SFE) models. With this approach, instead of setting their price bids (Bertrand) or quantities (Cournot), see \cite{microeconomics}, firms bid their choices of supply functions and the predicted outcome is a Nash equilibrium of the game. 
SFE models were first introduced by \cite{sfe}, and then applied to electricity markets by \cite{green}). While there is a vaste amount of literature directed to the study of SFE outcomes in uniform-price auctions 
(e.g., \cite{david}, \cite{baldick},  \cite{anderson}, \cite{linear_sfe}, \cite{holmberg2010}, \cite{correa2014pricing}), less clear is the behavior of SFE models when discriminatory prices are considered. In pay-as-bid auctions, firms overbid to ensure profit and their behavior is less predictable. In \cite{karaca2017game} and \cite{karaca2019designing}, the authors use tools from auction theory to propose an alternative mechanism based on Vickrey–Clarke–Groves (VCG) mechanism to incentivize
truthful bidding.

Our focus is on existence and characterization of Nash equilibria in supply functions with the pay-as-bid remuneration and asymmetric firms. We determine conditions on the strategy space under which existence is guaranteed and best responses can be characterized with piecewise affine functions. Our work is related to \cite{sfe_pab}, where uncertainty is considered.
The authors determine conditions on the hazard rate of the demand distribution to ensure existence of Nash equilibria which is in general not guaranteed. We instead study the problem from
a deterministic perspective with the objective of determining a rather tractable model. Although different in the purpose, it is relevant to mention \cite{genc}, where supply function equilibria game models are compared for uniform-price and pay-as-bid auctions. Our model differs from the one in \cite{genc} as they consider inelastic time-varying demand and single-step marginal cost functions.

The rest of the paper is organized as follows. In Section 2, we present the general setting and the model. In Section 3, we state and comment the main result. In Section \ref{ss:ex}, we show an example of
pay-as-bid auction game and we compare the Nash equilibrium
outcomes depending on the choice of the strategy space. In
Section \ref{sec:sm}, we discuss the supermodularity of the game and
anticipate some current work. Section 4 concludes the paper
and discusses some further research. 



\section{Model}
Throughout, $\R_+$ will stand for the set of nonnegative reals. For a non-empty interval $\mc I\subseteq\R$, we shall denote by $\mc C^0(\mc I)$ and $\mc C^k(\mc I)$, respectively, the sets of continuous and $k$-times continuously differentiable functions $f:\mc I\to\R$.  

\subsection{Problem setting}
We consider a system with:
\begin{itemize} 
	\item[-] an agent set $\N=\{1,\dots, n\}$  of $n$ firms equipped with \textit{cost functions} $C_i(q)=c_iq^2\,,$ with $c_i\geq0$ for $i$ in $\N$, where $q$ denotes the sold quantity;
	\item[-] an aggregate \textit{demand function} $D=N-\gamma p$ with $N\geq 0$ and $\gamma>0$, which returns the quantity $D(p)$ that consumers are willing to buy at a (maximum) unit price $p$. We define $\hat{p}$ as the price such that $D(\hat{p})=0$, i.e., $\hat{p}=\frac{N}{\gamma}$.
\end{itemize}

The \textit{strategy} of an agent $i$ in $\N$
is a supply function  
belonging to a predetermined nonempty subset $\mc A$ of the set of non-decreasing continuous functions that are $0$ in $0$, i.e.,
\begin{equation}\label{eq:F}
	\mc F=\{S\in \mc C^0([0, \hat{p}])\,,\,S_i(0)=0\,,\,S \emph{ non-decreasing}\}\,.
\end{equation}
The supply function $S_i$ returns the quantity $q=S_i(p)$ that the agent is willing to produce at (minimum) unit price $p$. 
The \textit{strategy configuration} of the game combines all strategies, that is, $\textbf{S}=(S_1,\dots, S_n)$. 
For an agent $i$ in $\N$ and a strategy configuration $\textbf{S}$, we shall refer to the other agents' strategies with $S_{-i} =\{S_j\}_{j \neq i}$. 

Given a demand function $D(p)$ and a strategy configuration $\textbf{S}$, the \textit{equilibrium marginal price} is determined as the price that matches total demand and total supply, that is, $p^*$ in $[0, \hat{p}]$  satisfying
\begin{equation} \label{eq:equilibrium}
	D(p^*)= \sum_{i=1}^nS_i(p^*)\,.
\end{equation}  	
We remark that existence and uniqueness of an equilibrium marginal price $p^*$ in $[0, \hat{p}]$ are guaranteed by the assumptions of a strictly decreasing continuous demand function and increasing continuous supply functions satisfying $S_i(0)=0$ for all $i$. 
The equilibrium marginal price determines the total quantity that will be sold by each agent in the auction, that is, $q^*_i = S_i(p^*)$ for every $i$ in $\N$. An example of equilibrium marginal price is depicted on the left of Fig.\ref{fig:paba}.

\subsection{Pay-as-bid auction game}

We define the following class of games based on the pay-as-bid remuneration. For a given $\mc A \subseteq \mc F$, the \textit{pay-as-bid} (PAB) \textit{auction} is a game with agent set $\N$, strategy space $\A$ and utilities, for every $i$ in $\N$,
\begin{equation}\label{eq:utility_pab}
	u_i(S_i, S_{-i}) :=  p^* S_i(p^*) - \int_0^{p^*}S _i(p)\dd p - C_i(S_i(p^*))\,,
\end{equation}	where $p^*:=p^*(S_i, S_{-i})$ is the equilibrium marginal price satisfying (\ref{eq:equilibrium}).
We shall denote the PAB auction game with $\mc U = (\mc N, \mc A, \{u_i\}_{i\in \mc N})$,		

\begin{figure}
	\centering
	\includegraphics[width=0.22\textwidth]{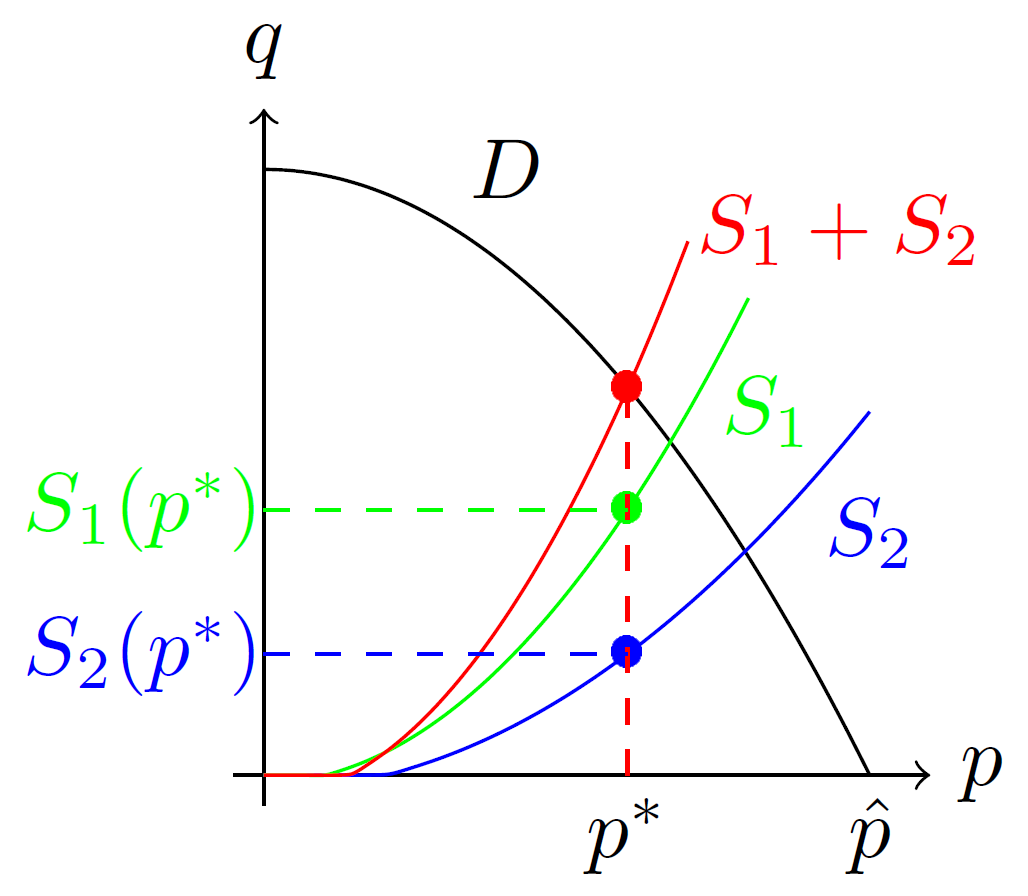}
	\includegraphics[width=0.22\textwidth]{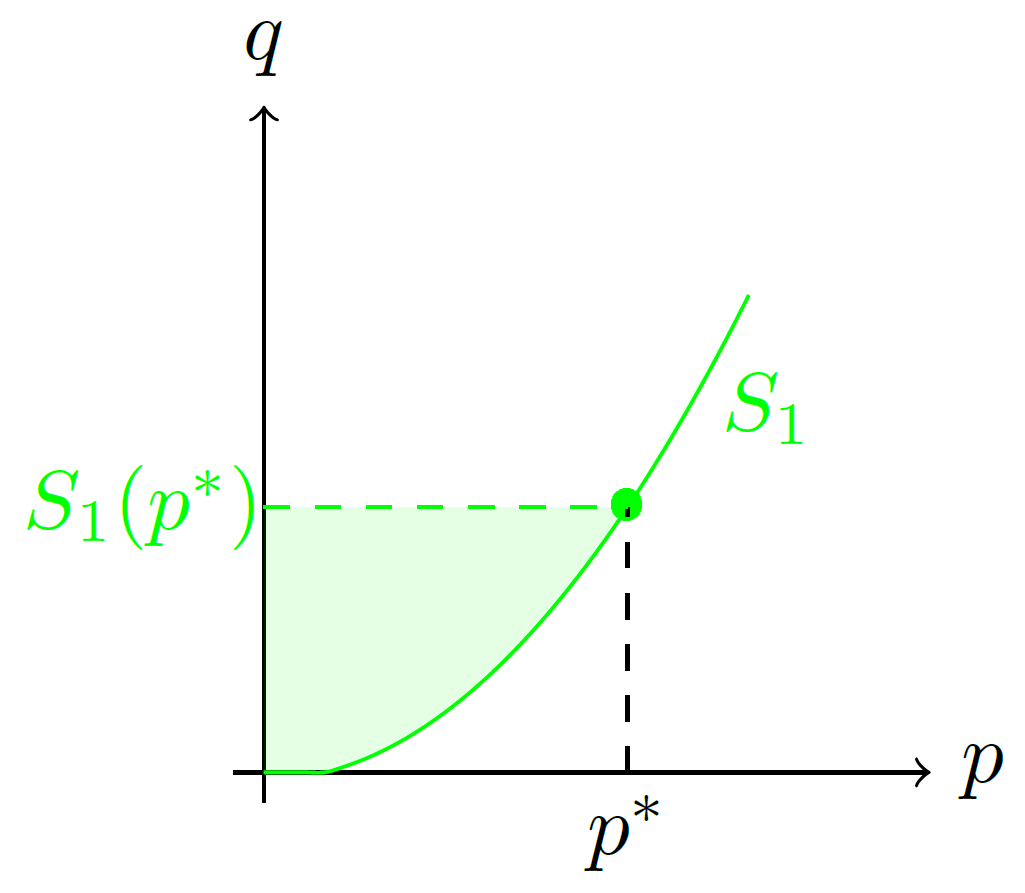}
	\caption{The equilibrium marginal price (on the left) and the pay-as-bid remuneration (on the right).}
	\label{fig:paba}
\end{figure}  
In words, agent $i$ sells $S_i(p^*)$ at the bid price and the final utility is given by the total revenue minus the production cost. Indeed, notice that, when $S_i$ is differentiable
$$
\int_0^{p^*}p\,S'_i(p)\dd p 
=   p^* S_i(p^*) - \int_0^{p^*}S _i(p)\dd p \,.
$$
Also, if $S_i$ is invertible, the total revenue in the PAB auction game equals the integral from 0 to $S_i(p^*)$ of the inverse of $S_i$, that is, the price function $P_i(q):=S^{-1}_i(q)$ of agent $i$. The price function assigns to each quantity the marginal price at which agents are willing to sell such quantity for. Therefore, its integral from $0$ to $q^*_i$ determines the total pay-as-bid remuneration for agent $i$ for a quantity $q_i^*$. By considering the formula in (\ref{eq:utility_pab}), we do not need to make any assumption on $S_i$. 

An example of remuneration of the PAB auction game is depicted on the right of Figure \ref{fig:paba}. 
When the supply function is $S_1$ and the equilibrium marginal price is $p^*$, the total revenue for agent $1$ coincides with the green area (the utility is then given by revenue minus costs).

Throughout the analysis, we shall focus on existence and characterization of Nash equilibria of the PAB auction game. A strategy configuration $\textbf{S}$ is a (pure strategy) \textit{Nash equilibrium} if, for every $i$ in $\N$, $S_i$ maximizes the utility given the other agents' strategies. Let $S_{-i}$ in $\A^{\N\setminus \{i\}}$. 
We shall refer to the set
\begin{equation}\label{eq:def_br}
	\mc B_i(S_{-i}) = \text{argmax}_{S_i \in \mc A} u_i(S_i, S_{-i})
\end{equation}
as the \textit{best response} of agent $i$ to $S_j$. Then, $\textbf{S}^*$ is a Nash equilibrium if and only if $S^*_i$ in $\mc B_i(S^*_{-i}) $ for every $i$ in $\N$.

Let us observe that, when the supply functions can be generic non-increasing continuous function, that is, when $\A=\mc F$ as in (\ref{eq:F}), the PAB auction game does not admit Nash equilibria in general. 
Let $S^0 \equiv 0$ denote the supply function that is zero in all the interval $[0, \hat{p}]$. Then, the set of Nash equilibria is either empty or equal to $S^*=\{S^0\}_{i\in \mc N}$. More precisely, we shall prove that the best-response is either $S^0$ or does not exist. We remark that the case when $S^*=\{S^0\}_{i\in \mc N}$ is Nash equilibrium is a limit case, which is not particularly interesting as all agents are selling a zero amount of quantity.
\begin{prop}\label{pr:no_br}
	Consider the PAB auction game with strategy space $\mc A= \mc F$ as in (\ref{eq:F}). Then, for every $i$ in $\N$ and $S_{-i}$ in $\mc A^{\N\setminus \{i\}}$, $\mc B_i(S_{-i}) = \emptyset$ or $\mc B_i(S_{-i}) = S^0$.
\end{prop} 
\begin{pf} See Appendix \ref{proof-prop-nobr}. \qed\end{pf}
\textit{Remark}
	The proof of Proposition \ref{pr:no_br} suggests that best responses would exist if one could use step functions. However, enlarging the strategy space to discontinuous functions would lead to a number of different technical difficulties.
	For instance, one has to solve some technical problems in the definition of the game. Indeed, the existence of a unique marginal equilibrium price $p^*$ as the unique solution of \eqref{eq:equilibrium} is not guaranteed anymore. 
	Anyway, even when we technically solve such problem, Nash equilibria might fail to exist.

Thus, we observed that, in the general settings, Nash equilibria might fail to exist. This gives the motivation for the $K$-Lipschitz assumption, which guarantees existence and characterization of Nash equilibria.


\section{Main result}
In this section, we state and comment the main result of the paper. 
Under the assumption of $K$-Lipschitz supply functions, we prove existence and characterization of Nash equilibria in the PAB auction game. We then illustrate and comment some examples and we discuss the supermodularity of the game.

As previously observed, one of the main issues is that, without any particular restriction on the strategy space, the best response is a step function and existence of Nash equilibria is not guaranteed. Indeed, the best response is an empty set when $\mc F=\mc A$.
We solve this problem by restricting the strategy space of the agents to the space of $K$-Lipschitz supply functions, for a fixed $K>0$. 

We recall that a function $S: [0, \hat{p}]\rightarrow [0, \infty)$ is $K$-\textit{Lipschitz} for $K>0$ if 
$$
|S (x)-S(y)|\leq K|x -y|\,,\quad \forall x ,y \in[0, \hat{p}] \,,\,x \neq y \,.
$$
Let us then define 	\begin{equation}\label{eq:strategy_space}
	\A_K:=\{S \in \mc F\,: S \text{ is}\textit{ K-Lipschitz} \}\,.
\end{equation}
By restricting the strategy space to $K$-Lipschitz supply functions, that is, with $\mc A = \mc A_K$, we can characterize best response functions and prove existence of Nash equilibria, as showed in the following result. 

\begin{thm}\label{th:pab_eq}
	Consider the PAB auction game with strategy space $\mc A =  \mc A_K$ as in \eqref{eq:strategy_space}.  Then, there exists at least one Nash equilibrium $\textbf{S}^*$ such that for every agent $i$ in $\mc N$: 
	\begin{equation}\label{eq:ch_br}
			S^*_i(p)=K[p-p_i]_+\,,
	\end{equation}
 for some $p_i $ in $[0, \hat{p}]$. 
\end{thm}

Theorem \ref{th:pab_eq} guarantees existence of Nash equilibria. Also, a characterization is given: in Nash equilibria, the strategies of the agents are piecewise affine functions with slope $K$. 

The analysis that led to the proof of Theorem \ref{th:pab_eq} is structured as follows. First, we prove that, if the strategy set of all agents is restricted to $K$-Lipschitz supply functions, not only best responses do exist, but it is rather simple to determine their structure. They are a subset of piecewise affine that can be parametrized by a single scalar value for every agent (see Remark 1). This in particular implies that Nash equilibria  of the original PAB auction game correspond to those of a finite dimensional game whereby the bidders have to choose such scalar parameter. Then, it is possible to define and study such finite-dimensional game, in particular showing that the utility functions are continuous and quasi-concave in such parameters and proving existence of Nash equilibria as a consequence (see Proposition 20.3 in \cite{osborne1994course}, or also \cite{debreu}, \cite{glicksberg}, and \cite{fan}).

\begin{figure}
	\centering
	\includegraphics[width=0.15\textwidth]{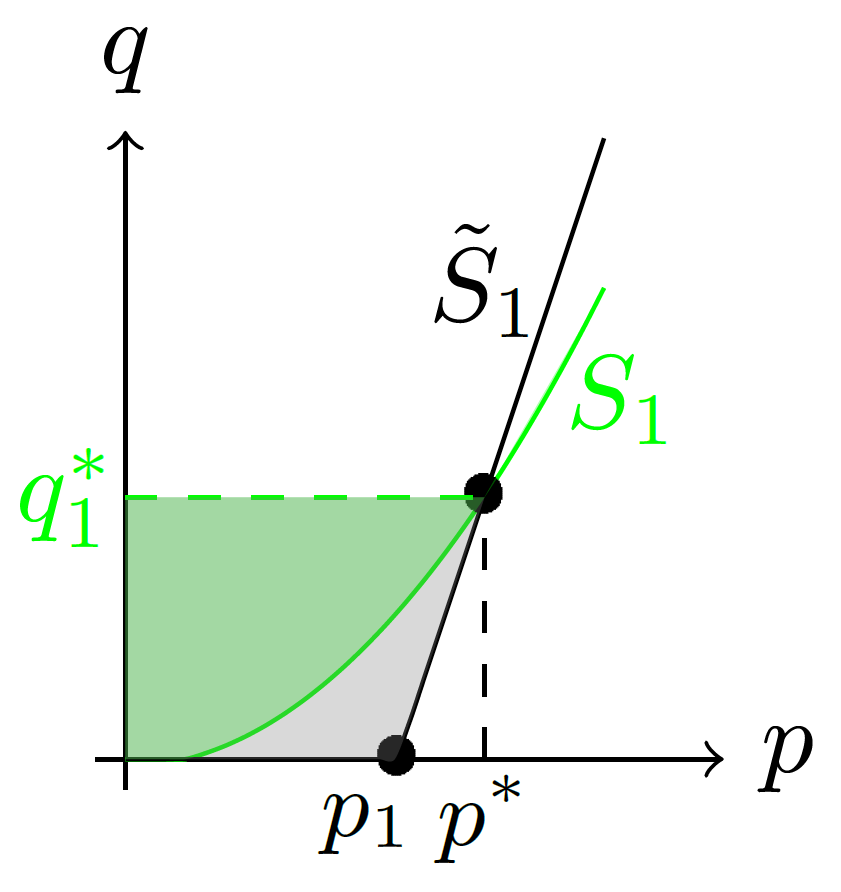}
	\includegraphics[width=0.15\textwidth]{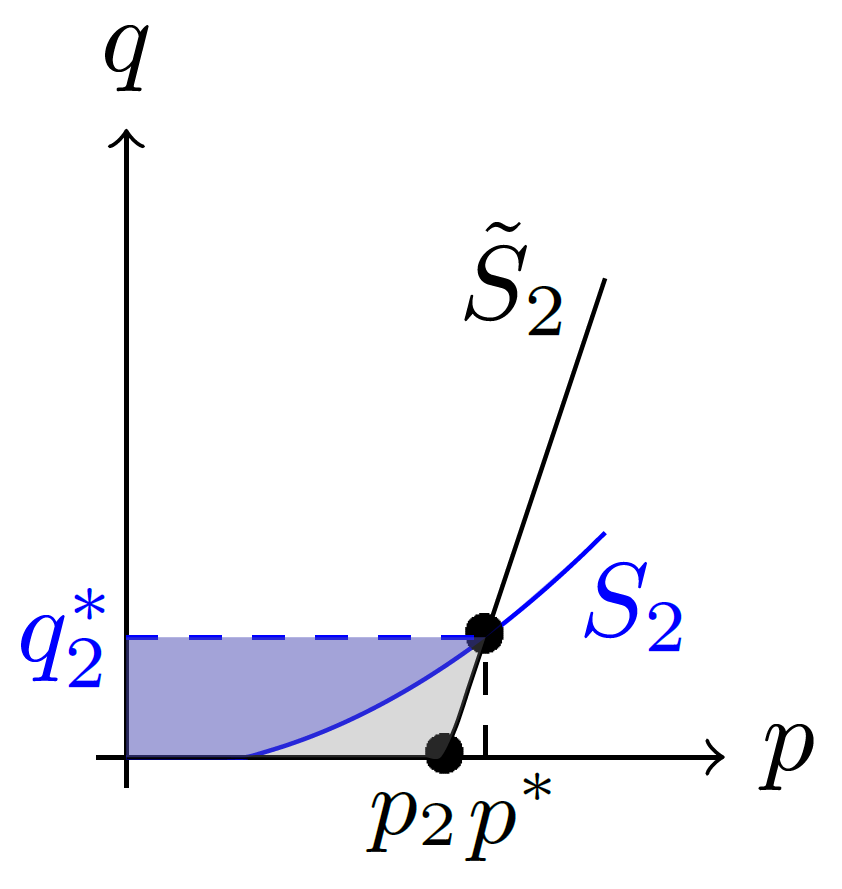}
	\includegraphics[width=0.15\textwidth]{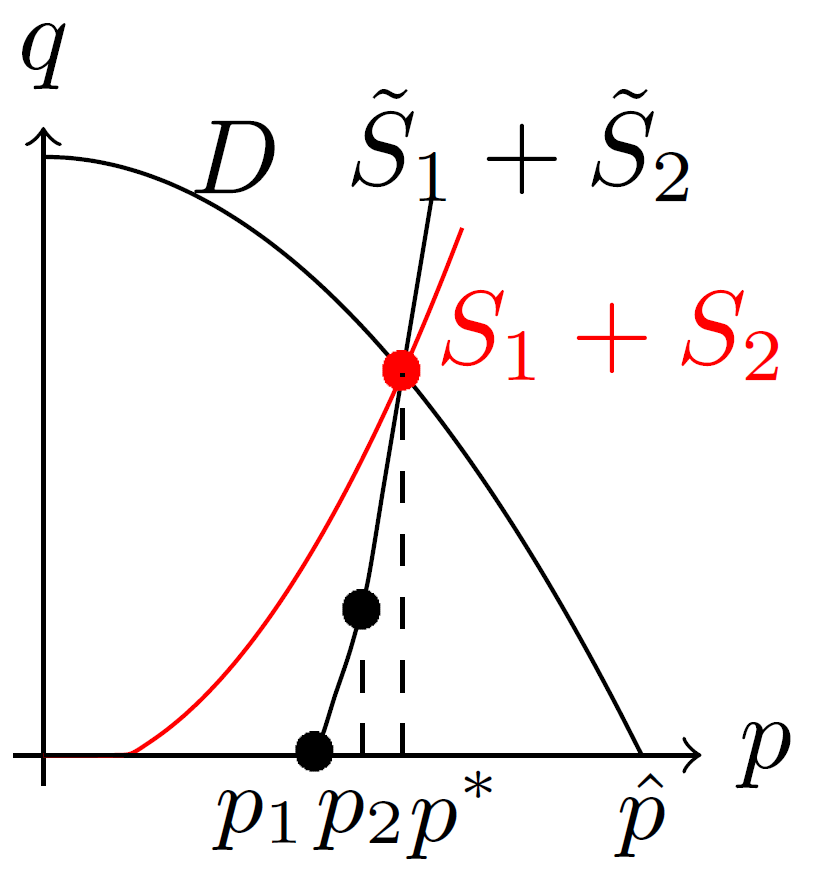}
	\caption{Idea behind characterization of best responses (see Remark 1).}
	\label{fig:Klip_sf}
\end{figure}  

\textit{Remark 1.}
	In Fig. \ref{fig:Klip_sf}, we illustrate the idea behind the characterization of best responses. Consider two generic supply functions $S_1$ and $S_2$ as in Fig.\ref{fig:paba}. Notice that when playing $\tilde{S}_1(p) = K[p-p_1]_+$ for $p_1$ as in figure, 
	agent $1$ receives a higher utility than the one obtained by playing $S_1$. Indeed, the remuneration increases (colored areas), while the equilibrium price does not change, thus yielding to the same sold quantity. The same happens for agent $2$ when playing $\tilde{S}_2(p)=K[p-p_2]_+$ instead of $S_2$. Then, for any strategy $S_i$, it is possible to construct another supply $\tilde{S}_i$ of the form in \eqref{eq:ch_br} yielding to a higher utility. Thus, best responses must have such form.

\subsection{Examples and discussion on the choice of the strategy space}\label{ss:ex}
In this section, we show an example of the PAB auction game and we compare Nash equilibria for different values of $K>0$.

Let us consider the following setting. There are $n=4$ agents partecipating in the auction game and their costs functions are: $$\begin{aligned}
	C_1(q)=\frac{1}{4}q^2\,,&\quad C_2(q)=\frac{1}{2}q^2\,,\\C_3(q)=q^2\,,&\quad C_4(q)=2q^2\,.\end{aligned}$$ The aggregate demand function is given by $D(p)=100-10p$ and, therefore, $\hat{p}=10$ (recall that, by definition, $D(\hat{p})=0$). 

\begin{figure}
	\centering	\includegraphics[width=0.45\textwidth]{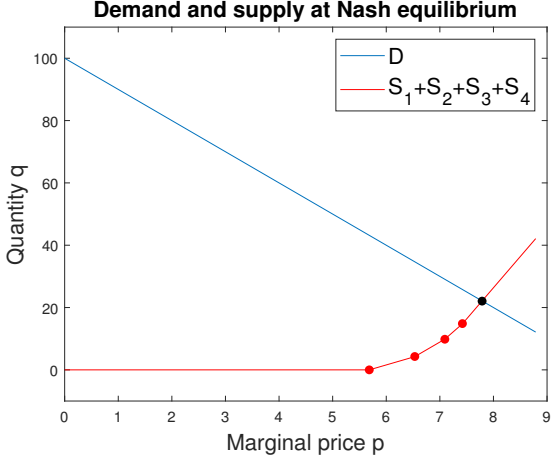}
	\caption{Aggregate demand and supply curves at Nash equilibrium for the setting in Section 4 and $K=5$.}
	\label{fig:ex}
\end{figure}

We now consider the PAB auction game with strategy space $\mc A_K$ for $K=5$. Theorem \ref{th:pab_eq} guarantees that there exists at least one Nash equilibrium $\textbf{S}^*=(S^*_1, S^*_2, S^*_3, S^*_4)$ of the form 
$$
\begin{aligned}
	S^*_1(p)=5[p-p_1]_+\quad &S^*_2(p)=5[p-p_2]_+\,\\
	S^*_3(p)=5[p-p_3]_+\quad &S^*_4(p)=5[p-p_4]_+\,
\end{aligned}
$$
for some $p_i \in [0, \hat{p}]$ with $i=1,\dots,4$. The configuration $\textbf{S}^*$ is indeed a Nash equilibrium for $p_1\approx5.68$, $p_2\approx6.53$, $p_3\approx7.09$ and $p_4\approx7.42$. In Figure \ref{fig:ex}, we see the aggregate demand and suppy when agents bid the Nash equilibrium supply functions. The equilibrium price is then $p^*\approx 7.79$ and the utilities are $u_1(S^*)\approx 43.2$, $u_2(S^*)\approx 25.25$, $u_3(S^*)= 13.78$ and $u_4(S^*)= 7.22$.       

\begin{figure}
	\centering	\includegraphics[width=0.45\textwidth]{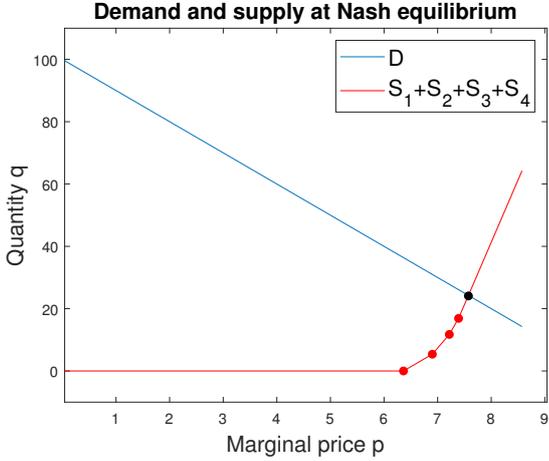}
	\caption{Aggregate demand and supply curves at Nash equilibrium for the setting in Section 4 and $K=10$.}
	\label{fig:ex2}
\end{figure}
Let us now study the PAB auction game in the same setting when $K=10$. In this case, we find the Nash equilibrium:
$$
\begin{aligned}
	S^*_1(p)=10[p-p_1]_+\quad &S^*_2(p)=10[p-p_2]_+\,\\
	S^*_3(p)=10[p-p_3]_+\quad &S^*_4(p)=10[p-p_4]_+\,
\end{aligned}
$$
with $p_1\approx6.36$, $p_2\approx6.9$, $p_3\approx7.22$ and $p_4\approx7.39$. In Figure \ref{fig:ex2}, we see the aggregate demand and suppy when agents bid the Nash equilibrium supply functions. The equilibrium price, in this second case, is $p^*\approx 7.57$ while the utilities are $u_1(S^*)\approx 47.74$, $u_2(S^*)\approx26.07$, $u_3(S^*)\approx 13.66$ and $u_4(S^*)\approx 7$.

Notice that, with a higher value of $K$, the equilibrium price decreases, while utilities increase for firms with lower costs and decrease for firms with higher costs. This observation leads to a discussion on the value of $K$. Indeed, our model requires $K$ to be fixed, but for any $K$ we obtain different equilibrium outcomes.  Recall that, according to Proposition \ref{pr:no_br} and the following remark, best responses would exist in the general setting if one could use step functions. Also, as $K$ increases, we are enlarging the strategy space. Therefore, a fundamental example that requires a deeper study is the case when $K$ approaches infinity. 

For instance, let us consider $K=1000$. In this case, we find the Nash equilibrium:
$$
\begin{aligned}
	S^*_1(p)=1000[p-p_1]_+\quad &S^*_2(p)=1000[p-p_2]_+\,\\
	S^*_3(p)=1000[p-p_3]_+\quad &S^*_4(p)=1000[p-p_4]_+\,
\end{aligned}
$$
with $p_1\approx 7.261$, $p_2\approx7.269$, $p_3\approx7.272$ and $p_4\approx7.274$. In Figure \ref{fig:ex2}, we see the aggregate demand and suppy when agents bid the Nash equilibrium supply functions. The equilibrium price, in this second case, is $p^*\approx 7.276$ while the utilities are $u_1(S^*)\approx 52.84$, $u_2(S^*)\approx26.45$, $u_3(S^*)\approx 13.23$ and $u_4(S^*)\approx 6.62$. 

We can observe that, as $K$ increases, all $p_i$ approach the equilibrium price $p^*$, while utilities are different among agents (due to the heterogenous costs). Current work includes a characterization of Nash equilibria that could lead to a comparison between our model and the classic economic models, such as Bertrand and Cournot models. Our conjecture is that, as $K$ goes to infinity, the equilibrium outcome of our model will approach the Bertrand equilibrium, when it exists. Indeed, in the limit case, agents will basically bid the equilibrium price. On the other hand, for lower values of $K$, we could observe similarities between our model and the Cournot model. The optimal value of $K$ shall be set by the auctioneer depending on the desired equilibrium outcome.

\begin{figure}
	\centering	\includegraphics[width=0.45\textwidth]{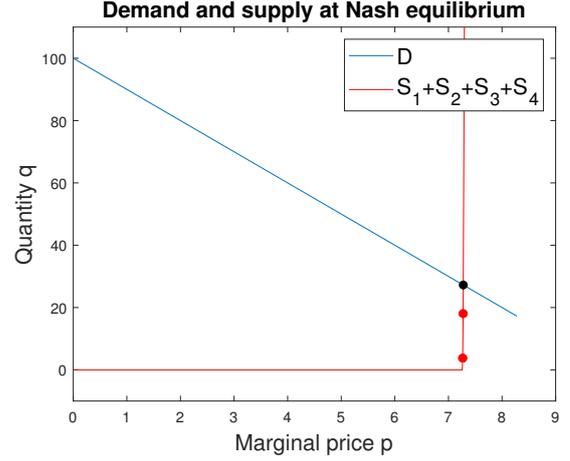}
	\caption{Aggregate demand and supply curves at Nash equilibrium for the setting in Section 4 and $K=1000$.}
	\label{fig:ex3}
\end{figure}

	\subsection{Discussion on supermodularity}\label{sec:sm}
Supermodular games (\cite{Topkins:1979,Vives:1990}; \cite{Topkins:1998,Milgrom.ROberts:1990}) are an important class of games. They are characterized by “strategic complementarities”, that is, when one agent increases the strategy, the others have an incentive in doing the same. In the previous sections, we observed that, in order to find Nash equilibria, we can restrict the strategy space to considering just functions as in \eqref{eq:ch_br}, which are parametrized by just one parameter, that is, $p_i \in[0, \hat{p}]$, for $i $ in $\N$. Below we recall the definition of supermodular games and we briefly discuss the properties of the restricted game in this setting. First, we formally define the restricted game  $\mc U_r$ and  we observe that it is not supermodular in general. Anyway, our conjecture is that, when the demand is linear, the game is somehow "piece-wice supermodular". Current work includes a deeper analysis in this direction that could allow to exploit the fundamental properties of supermodular games. 

Let $x, y$ in $\R^{\mc I}$ for some finite set $\mc I$. In the following, we shall consider the component-wise partial order $\leq$, formally
defined by
$$
x \leq y \Leftrightarrow x_i \leq y_i \quad \forall i\in \mc N\,.
$$
A game $(\mc N, \mc A, \{u_i\}_{i\in \mc N})$ is \textit{supermodular} if for all $i\in \mc N$
\begin{enumerate}
	\item $\mc A$ is a compact set of $\R$;
	\item $u_i$ is upper semi-continuous in $\mc A$ and $\mc A^{\mc N\setminus i}$;
	\item $u_i$ has increasing differences in $(p_i,p_{-i})$, namely, if for all $p_i'\geq p_i$ and $p_{-i}'\geq p_{-i}$ it holds
	\begin{equation}\label{eq:incr_diff}
		\hspace{-0.3cm}	u_i(p_i',p_{-i}')-u_i(p_i,p_{-i}')\geq u_i(p_i',p_{-i})-u_i(p_i,p_{-i})\,.
	\end{equation}
\end{enumerate}

The restricted game $\mc U_r = (\mc N, \mc A_r, \{u^r_i\}_{i\in \N})$ has finite-dimensional strategy space $\mc A_r=[0, \hat{p}]$ and utilities, for $i$ in $\N$,
\begin{equation}\label{eq:u_r}
	\begin{aligned}
		u^r_i(p_i, p_{-i}) := &\,p^*K[p^*-p_i]_+-\frac{\left(K[p^*-p_i]_+\right)^2}{2K}\\ &\,-C_i\left(K[p^*-p_i]_+\right)	 \\
		\text{s.t.: }& \,\, 	 D(p^*)=\sum_{i=1}^nK[p^*-p_i]_+\,,
	\end{aligned}
\end{equation}
where  $p_{-i}=\{p_j\}_{j\neq i}$ denotes the actions of all the remaining players. Let us first observe that conditions (1) and (2) are satisfied for $\mc A_r =[0, \hat{p}]$ and $u_i^r$ in (\ref{eq:u_r}).
In other words, the game $\mc U_r$ is super-modular if the marginal
utility $u^r_i(p_i',p_{-i})-u^r_i(p_i, p_{-i})$ of every agent $i$ is a monotone nondecreasing function of the strategy profile $p_{-i}$ of the other agents. 

Supermodularity of the restricted game $\mc U_r$ is not guaranteed in the general case, as shown in the following example.
		{\textit{Example 1.} Let us consider a game with two agents, that is, $n=2$, and let $D(p)=N-\gamma p$ with $N=100$ and $\gamma =1$. Let us also assume that agent $1$ has quadratic costs, that is, $C_1(q)=\frac{1}{2}c_1q^2$ with $c_1=1$. We shall discuss the supermodular property of the restricted game $\mc U_r$ with $K=1$. 
			
			Let $p_2 =0$ and $p_1 = \frac{N}{\gamma+1}=50$. Observe that, for such values of $p_1$ and $p_2$, we find
			$$
			N-\gamma p^* = [p^*-p_1]_+ +[p^*-p_2]_+ \Leftrightarrow p^*= \frac{N}{\gamma +1}=50\,.
			$$ 
			Therefore, agent $1$ does not sell any quantity and her utility is $u_1(p_1,p_2)=-C_1(0)=0\,.$ The same holds if she increases her strategy. For instance, for $p_1'=50.2$, we find $u_1(p'_1,p_2)=-C_1(0)=0\,.$ 
			
			On the other hand, the utility of agent $1$ changes when agent $2$ increases her strategy. Let $p_2'=1$. Then, we find
			$$
			\begin{aligned}
				N-\gamma p^*&= [p^*-p_1]_+ +[p^*-p_2']_+ \\&\Leftrightarrow \quad p^*= \frac{N+p_1+p_2'}{\gamma +2} = 50.3\bar{3}
			\end{aligned}
			$$
			Observe that $p^*>p_1$ and therefore agent $1$ sells a quantity $p^*-p_1$. The utility is then given by
			$$
			u_1(p_1, p_2') = \frac{(p^*)^2}{2}- \frac{(p_1)^2}{2}-\frac{c_1}{2}(p^*-p_1)^2 = 50/3\,.
			$$
			Similarly, for $p_1'=50.2$ and $p_2'=1$, we find $p^*\approx 50.4$ and $u_1(p_1', p_2')\approx 10.04$.
			
			The game is supermodular if the utilities $u_1$ and $u_2$ satisfy (\ref{eq:incr_diff}) for all $p_1'\geq p_1$ and $p_2'\geq p_2$. According to our previous computations, for $ p_1'=50.2\geq p_1=50$ and $p_2'=1\geq p_2=0$, we obtain
			$$
			\begin{aligned}
				u_1(p_1',p_2')-u_1(p_1,p_2')&\approx-6.63\\&\ngeq u_1(p_1',p_2)-u_1(p_1,p_2)=0\,.
			\end{aligned}
			$$
			Therefore, the game is not supermodular.}
		
		Example 1 shows that the restricted game $\mc U_r$ can fail to be supermodular also in the case with two agents, affine demand and quadratic costs.  Anyway, we observed that, if the demand is affine, the restricted utility $u^r_i$ of an agent $i$ in $\mc N$ satisfy (\ref{eq:incr_diff}) in some intervals. More precisely, we need to require that, for all possible combinations of $p_i, p'_i,p_{-i},p_{-i}'$: 
		\begin{enumerate}
\item agent $i$  always sells a non-zero quantity, and
\item the number of agents selling a non-zero quantity in the game remains constant.
		\end{enumerate}
	Current work includes a precise definition of the piecewise supermodularity. We aim to exploit this observation for the the study of uniqueness of Nash equilibria. Also, we intent to investigate conditions for the convergence to Nash equilibria in this setting.

\section{Conclusion}
We have studied a supply function equilibrium model with pay-as-bid remuneration and asymmetric firms. We have proved existence of (pure strategy) Nash equilibria when the strategy space of the agents is restricted to the space of $K$-Lipschitz supply functions. In this setting, a characterization of Nash equilibria is given: strategies at equilibrium take the form of stepwise affine functions with slope $K$.

In Section \ref{ss:ex}, we discuss with some examples the choice of the strategy space, i.e., the choice of the parameter $K$. Current work includes a characterization of Nash equilibria and a comparative statics for the special case when $K$ approaches infinity. In Section \ref{sec:sm}, the supermodular property of the game is briefly discussed. The game is not supermodular in general, although we observed that, when the demand is affine, the restricted game is somehow "piece-wice" supermodular. Current work includes a deeper analysis in this direction. Our conjecture is that convergence to Nash equilibria is guaranteed when the demand is affine. 

Further work comprehends conditions for uniqueness of Nash equilibria. Also, we intend to include uncertainty in our model. Motivated by the current structure of electricity markets, we aim to study the concatenation of a uniform-price auction and a pay-as-bid one, modeled as a two-stage game.
\bibliography{bib} 

\begin{thebibliography}{28}
\providecommand{\natexlab}[1]{#1}
\providecommand{\url}[1]{\texttt{#1}}
\providecommand{\urlprefix}{URL }
\expandafter\ifx\csname urlstyle\endcsname\relax
  \providecommand{\doi}[1]{doi:\discretionary{}{}{}#1}\else
  \providecommand{\doi}{doi:\discretionary{}{}{}\begingroup
  \urlstyle{rm}\Url}\fi

\bibitem[{Anderson and Philpott(2002)}]{anderson}
Anderson, E.J. and Philpott, A.B. (2002).
\newblock Using supply functions for offering generation into an electricity
  market.
\newblock \emph{Operations research}, 50(3), 477--489.

\bibitem[{Baldick et~al.(2004)Baldick, Grant, and Kahn}]{linear_sfe}
Baldick, R., Grant, R., and Kahn, E. (2004).
\newblock {Theory and Application of Linear Supply Function Equilibrium in
  Electricity Markets}.
\newblock \emph{Journal of Regulatory Economics}, 25(2), 143--167.

\bibitem[{Baldick and Hogan(2001)}]{baldick}
Baldick, R. and Hogan, W. (2001).
\newblock Capacity constrained supply function equilibrium models of
  electricity markets: Stability, nondecreasing constraints, and function space
  iterations.

\bibitem[{Correa et~al.(2014)Correa, Figueroa, Lederman, and
  Stier-Moses}]{correa2014pricing}
Correa, J.R., Figueroa, N., Lederman, R., and Stier-Moses, N.E. (2014).
\newblock Pricing with markups in industries with increasing marginal costs.
\newblock \emph{Mathematical Programming}, 146(1), 143--184.

\bibitem[{David(1993)}]{david}
David, A.K. (1993).
\newblock Competitive bidding in electricity supply.
\newblock In \emph{IEE proceedings C-Generation, transmission and
  distribution}, volume 140, 421--426. IET.

\bibitem[{Debreu(1952)}]{debreu}
Debreu, G. (1952).
\newblock A social equilibrium existence theorem.
\newblock \emph{Proceedings of the National Academy of Sciences}, 38(10),
  886--893.

\bibitem[{Fabra et~al.(2006)Fabra, von~der Fehr, and Harbord}]{fabra}
Fabra, N., von~der Fehr, N.H., and Harbord, D. (2006).
\newblock Designing electricity auctions.
\newblock \emph{The RAND Journal of Economics}, 37(1), 23--46.

\bibitem[{Fan(1952)}]{fan}
Fan, K. (1952).
\newblock Fixed-point and minimax theorems in locally convex topological linear
  spaces.
\newblock \emph{Proceedings of the National Academy of Sciences of the United
  States of America}, 38(2), 121.

\bibitem[{Genc(2009)}]{genc}
Genc, T.S. (2009).
\newblock Discriminatory versus uniform-price electricity auctions with supply
  function equilibrium.
\newblock \emph{Journal of optimization theory and applications}, 140(1),
  9--31.

\bibitem[{Glicksberg(1952)}]{glicksberg}
Glicksberg, I.L. (1952).
\newblock A further generalization of the {K}akutani fixed point theorem, with
  application to {N}ash equilibrium points.
\newblock \emph{Proceedings of the American Mathematical Society}, 3(1),
  170--174.

\bibitem[{Green and Newbery(1992)}]{green}
Green, R. and Newbery, D.M. (1992).
\newblock Competition in the british electricity spot market.
\newblock \emph{Journal of Political Economy}, 100(5), 929--53.

\bibitem[{Holmberg and Newbery(2010)}]{holmberg2010}
Holmberg, P. and Newbery, D. (2010).
\newblock The supply function equilibrium and its policy implications for
  wholesale electricity auctions.
\newblock \emph{Utilities Policy}, 18(4), 209--226.

\bibitem[{Holmberg(2009)}]{sfe_pab}
Holmberg, P. (2009).
\newblock Supply function equilibria of pay-as-bid auctions.
\newblock \emph{Journal of Regulatory Economics}, 36, 154--177.
\newblock \doi{10.1007/s11149-009-9091-6}.

\bibitem[{Kamgarpour(2018)}]{maryam}
Kamgarpour, M. (2018).
\newblock Game-theoretic models in energy systems and control.
\newblock DTU Summer School, Modern Optimization in Energy Systems.

\bibitem[{Karaca and Kamgarpour(2017)}]{karaca2017game}
Karaca, O. and Kamgarpour, M. (2017).
\newblock Game theoretic analysis of electricity market auction mechanisms.
\newblock In \emph{2017 IEEE 56th Annual Conference on Decision and Control
  (CDC)}, 6211--6216. IEEE.

\bibitem[{Karaca et~al.(2019)Karaca, Sessa, Walton, and
  Kamgarpour}]{karaca2019designing}
Karaca, O., Sessa, P.G., Walton, N., and Kamgarpour, M. (2019).
\newblock Designing coalition-proof reverse auctions over continuous goods.
\newblock \emph{IEEE Transactions on Automatic Control}, 64(11), 4803--4810.

\bibitem[{Klemperer and Meyer(1989)}]{sfe}
Klemperer, P. and Meyer, M. (1989).
\newblock Supply function equilibria in oligopoly under uncertainty.
\newblock \emph{Econometrica}, 57(6), 1243--77.

\bibitem[{Mas-Colell et~al.(1995)Mas-Colell, Whinston, and
  Green}]{microeconomics}
Mas-Colell, A., Whinston, M.D., and Green, J.R. (1995).
\newblock \emph{{Microeconomic Theory}}.
\newblock Number 9780195102680 in OUP Catalogue. Oxford University Press.

\bibitem[{Milgrom and Roberts(1990)}]{Milgrom.ROberts:1990}
Milgrom, P. and Roberts, J. (1990).
\newblock Rationalizability, learning, and equilibrium in games with strategic
  complementarities.
\newblock \emph{Econometrica}, 58(6), 1255--1277.
\newblock \urlprefix\url{https://doi.org/10.2307/2938316}.

\bibitem[{Osborne and Rubinstein(1994)}]{osborne1994course}
Osborne, M.J. and Rubinstein, A. (1994).
\newblock \emph{A course in game theory}.
\newblock MIT press.

\bibitem[{Paccagnan et~al.(2016)Paccagnan, Kamgarpour, and
  Lygeros}]{paccagnan2016aggregative}
Paccagnan, D., Kamgarpour, M., and Lygeros, J. (2016).
\newblock On aggregative and mean field games with applications to electricity
  markets.
\newblock In \emph{2016 European Control Conference (ECC)}, 196--201. IEEE.

\bibitem[{Rassenti et~al.(2003)Rassenti, Smith, and Wilson}]{rassenti}
Rassenti, S.J., Smith, V.L., and Wilson, B.J. (2003).
\newblock Discriminatory price auctions in electricity markets: low volatility
  at the expense of high price levels.
\newblock \emph{Journal of regulatory Economics}, 23(2), 109--123.

\bibitem[{Roozbehani et~al.(2010)Roozbehani, Dahleh, and
  Mitter}]{roozbehani2010stability}
Roozbehani, M., Dahleh, M., and Mitter, S. (2010).
\newblock On the stability of wholesale electricity markets under real-time
  pricing.
\newblock In \emph{49th IEEE Conference on Decision and Control (CDC)},
  1911--1918. IEEE.

\bibitem[{Tang et~al.(2016)Tang, Rajagopal, Poolla, and
  Varaiya}]{tang2016model}
Tang, W., Rajagopal, R., Poolla, K., and Varaiya, P. (2016).
\newblock Model and data analysis of two-settlement electricity market with
  virtual bidding.
\newblock In \emph{2016 IEEE 55th Conference on Decision and Control (CDC)},
  6645--6650. IEEE.

\bibitem[{Topkins(1979)}]{Topkins:1979}
Topkins, D.M. (1979).
\newblock Equilibrium points in nonzero-sum n-person submodular games.
\newblock \emph{SIAM Journal on Control and Optimization}, 17(6), 773--787.
\newblock \urlprefix\url{https://doi.org/10.1137/0317054}.

\bibitem[{Topkins(1998)}]{Topkins:1998}
Topkins, D.M. (1998).
\newblock \emph{Supermodularity and Complementarity}.
\newblock Princeton University Press.

\bibitem[{Ventosa et~al.(2005)Ventosa, Ba{\i}llo, Ramos, and Rivier}]{ventosa}
Ventosa, M., Ba{\i}llo, A., Ramos, A., and Rivier, M. (2005).
\newblock Electricity market modeling trends.
\newblock \emph{Energy policy}, 33(7), 897--913.

\bibitem[{Vives(1990)}]{Vives:1990}
Vives, X. (1990).
\newblock Nash equilibrium with strategic complementarities.
\newblock \emph{Journal of Mathematical Economics}, 19, 305--321.
\newblock \urlprefix\url{https://doi.org/10.1016/0304-4068(90)90005-T}.

\end{thebibliography}
\bibliographystyle{ieeetr}

\appendix
\section{}\label{proof-prop-nobr}
\begin{pf}	We shall prove that for every feasible supply function $S_i\neq S^0$ in $\mc A$, there exists another feasible supply function $\tilde{S}_i$ in $\mc A$ yielding to the same equilibrium price and a higher utility. Formally, let $S_i\neq S^0$ be any non-decreasing continuous function yielding to an equilibrium price $p^*\neq 0$. We shall then define $$\tilde{S}_i(p) := S_i\left(\frac{p^2}{p^*}\right)\,.$$ Observe that $\tilde{S}_i(0)=S_i(0)$ and $\tilde{S}_i(p^*)=S_i(p^*)$. Also $S_i(p)\geq \tilde{S}_i(p)$ for all $p $ in $[0, p^*]$. More precisely, we have that $S_i(p)= \tilde{S}_i(p)$ for all $p\in [0, p^*]$ if and only if $S_i$ is constantly equal to $0$. Therefore, if $S_i(p)\neq 0$ for some $p\in(0,\hat{p})$, then there must exist $p_0\in(0, p^*)$ such that $S_i(p_0)> \tilde{S}_i(p_0)$. For continuity, this implies that $\exists \epsilon >0$ such that $S_i(p)> \tilde{S}_i(p)$ for all $p \in (p_0-\epsilon, p_0+\epsilon)$.  We then obtain that
	$$
	\int_0^{p^*}\tilde{S} _i(p)\dd p<	\int_0^{p^*}S _i(p)\dd p\,,
	$$
	while the other terms remain constant. Consequently, we find that $u_i(\tilde{S}_i, S_{-i})> u_i(S_i, S_{-i})\,.$
	Then, the best response is either $S^0$ or does not exist. This concludes the proof.
\end{pf}
\end{document}